\documentclass[preprint, aps, draft]{revtex4}
\usepackage{amsfonts}
\usepackage{amssymb}
\usepackage{graphics}
\usepackage{epsfig}

\newcommand{\be}{\begin{equation}}
\newcommand{\ee}{\end{equation}}
\newcommand{\er}{\end{eqnarray}}
\newcommand{\ea}{\end{eqnarray}}
\newcommand{\br}{\begin{eqnarray}}
\newcommand{\ba}{\begin{eqnarray}}

\begin{document}

\title{Duality in nonlinear B$\wedge$F models: equivalence between self-dual and topologically massive Born-Infeld
B$\wedge$F models}

\author{R. Menezes${}^{b}$, J. R. S. Nascimento${}^{b}$ R. F. Ribeiro${}^{b}$ and C. Wotzasek${}^{a}$}
\email{rms, jroberto, rfreire@fisica.ufpb.br ; clovis@if.ufrj.br}
\affiliation{\em ${}^{a}$Instituto de F\'\i sica, Universidade Federal do Rio de Janeiro\\
21945-970, Rio de Janeiro, Brazil,\\
${}^{b}$Departamento de F\'\i sica, Universidade Federal da Para\'\i ba\\
58051-970, Jo\~ao Pessoa, Brazil}

\date{\today}

\begin{abstract}
We study the dual equivalence between the nonlinear generalization of the
self-dual ($NSD_{B\wedge F}$) and the topologically massive $B\wedge F$ models with particular emphasis on
the nonlinear electrodynamics proposed by Born and Infeld. This is done through
a dynamical gauge embedding of the nonlinear self-dual model yielding to a gauge invariant and dynamically
equivalent theory.
We clearly show that  nonpolinomial $NSD_{B\wedge F}$ models can be mapped, through a properly defined
duality transformation, into $TM_{B\wedge F}$ actions. The general result obtained is 
then particularized for a number of examples, including the 
Born-Infeld-BF (BIBF) model that has experienced a revival in the recent literature.
\end{abstract}

\maketitle

\section{INTRODUCTION}

This work is devoted to the study of duality symmetry in the nonlinear electrodynamics context with the presence of
a topological $B \wedge F$ term with particular emphasis to the Born--Infeld (BI) theory \cite{BI}.
These are models presenting a topological, first-order derivative coupling between forms of different ranks.
We investigate the existence of a constraint of self--duality in the massive, non invariant theory
($NSD_{B\wedge F}$) that is an extension of the model proposed in \cite{khare} in a different context.
To establish the duality mapping we adopt a new dynamical embedding formalism \cite{IW,AJRRW}, that is alternative to
the master Lagrangian approach \cite{DJ}, to obtain the gauge invariant $B\wedge F$ model.
This approach is also alternative to the idea of constraint conversion from the second to first-class constraints that characterizes the mapping from the non-invariant SD version into the gauge invariant version.
Our study also includes the case of
dynamical fermionic matter minimally coupled to the self-dual sector \cite{USP}.

This manifest realization of the duality mapping is important. The proof of duality, i.e., the equivalent description of a
physical phenomenon by distinct theories, is usually a non trivial task.
Such a dual description is desirable since it is important, in some instances, to have explicit symmetries manifest by
a redundant set of fields while in other cases, for instance during the process of canonical quantization, it is
desirable to work with a minimally complete set of variables.
To stablish the duality mapping, in the context of nonlinear $B\wedge F$ models, is a new and important result.

To stress the importance of nonlinear electrodynamics is almost unnecessary.
In fact, driven by the fact that nonlinear theories appear as
effective actions at different levels of String/M-theory, the Born-Infeld
nonlinear electrodynamics has observed an increasing revival in recent years.
The BI theory, an action for a bounded field strength, was proposed in the 30's,
as a nonlinear version of Maxwell electrodynamics,
in order to obtain a finite energy model for the electron.
The BI theory also arises as part of the low energy effective action of
the open superstring theory \cite{RMT85}.
A striking feature of BI theory, 
is that it admits BIon solutions, i.e., exact solutions of the full
non-linear theory with finite total
energy that can now be understood in terms of strings ending
on D$p$-branes, i.e., solitons of string theory
described by Dirac-Born-Infeld like actions \cite{L89,CM98,G98,G,CM}.
Supersymmetric extensions \cite{GNSS,BHL} and non-Abelian generalizations \cite
{T97,P99} of these nonlinear theories have also been constructed.
More recently, this string approach to nonlinear electrodynamics has been used
in the context of $AdS$/CFT correspondence (cf.~\cite{M99} for a review), to
obtain solutions describing baryon configurations which are consistent with
confinement \cite{C99}.
It is remarkable that recent works on open string
states in String/M-Theory has profited from insights
afforded by the BI approximation while,
in return, String/M-Theory has provided a rationale
for some of the, up to then, either mysterious or only partially understood
properties of this outstanding theory.

Other important results have been obtained in different frameworks. It has inspired
the formulation of other models, such as the Born-Infeld-Skyrmions, where
nonlinear terms are essential in order to obtain stationary solutions \cite
{DMS96,NdOO99}. Besides, both Maxwell and Born-Infeld
theories are singled out among all electromagnetic theories since they bear
both dual invariance \cite{GR95,GR96} and ``good propagations'' (in the
sense that excitations propagate without shocks) \cite{P68,DMS99}. It is
also remarkable that nonlinear electrodynamics satisfy the zeroth
and first laws of black hole mechanics \cite{R97}.

In this paper we are interested in a less explored application of
nonlinear electrodynamics, namely, the duality equivalence between
different models describing the same physical phenomenon, keeping invariant some properties such as
the number of degrees of freedom, propagator and equations of motion.
We define duality in a derivative sense \cite{BF} leading naturally to self and anti self dual solutions.
The paradigm of this equivalence \cite{DJ} is the well known duality between the
SD \cite{TPN} and MCS \cite{MCS} models in 2+1 dimensions.  This is
possible due to the presence of the topological and gauge
invariant Chern-Simons term (CST) \cite{CS} which is responsible
for fundamental features manifested by three-dimensional field
theories, such as parity breaking and anomalous spin \cite{??-01}.
The investigation of duality equivalence in three dimensions involving 
CST has had a long and fruitful history, beginning when Deser and Jackiw 
used the master action concept to prove the dynamical equivalence between 
the SD  and MCS theories \cite{DJ}, in this way proving the existence of 
a hidden symmetry in the SD version.
This approach has been extensively 
used thereafter, providing an invaluable tool in the study of the planar 
physics phenomena and in the extension of the bosonization program from 
two to three dimensions with important phenomenological consequences 
\cite{varios}.

The idea of including a topological term to produce non-trivial
phenomena has also been successful in $D > 3$.
In arbitrary dimensions duality will relate tensors of different ranks and $D=3$ is a
special case where vectors are dualized into vectors.
In particular for $D=4$  the
 inclusion of the so called $B \wedge F$ term has been responsible for new
and interesting features such as topological mass generation \cite{syracuse}
and statistical  transmutation \cite{indiano}.
In this respect we have examined recently \cite{BF}, using the gauge embedding procedure,
the duality between a gauge non invariant $B \wedge F$  model,
\begin{equation}
\label{PB10}
{\cal L}_{SD_{B \wedge F}} = \frac 12 m^2 A_\mu A^\mu - \frac 14 B_{\mu\nu} B^{\mu\nu}
+\frac {\chi\,\theta}4 \epsilon^{\mu\nu\lambda\rho} B_{\mu\nu} F_{\lambda\rho} \; ,
\end{equation}
 presenting the self-duality property and dual equivalence to the topologically massive $TM_{B \wedge F}$ model
\begin{equation}
\label{BI40}
{\mbox{}^*\cal L}_{TM_{B \wedge F}} =  \frac 1{12 m^2} \:H_{\mu\nu\lambda} H^{\mu\nu\lambda}  - 
\frac 14\: F_{\mu\nu} F^{\mu\nu}  - \frac \chi{2\,\theta} \:\epsilon^{\mu\nu\alpha\beta}\:
 A_\mu \partial_\nu B_{\alpha\beta}\; ,
\end{equation}
where $m$ is seem from the equations of motion to be the mass of the excitations. $A_\mu$ is a Maxwell-like vector
field and $B_{\mu \nu}$ is a rank-2, totally
 anti-symmetric Kalb-Ramond potential, whose field strengths read 
\begin{eqnarray}
\label{PB20}
F_{\mu\nu} &=& \partial_\mu A_\nu - \partial_\nu A_\mu \, , \nonumber\\
H_{\mu\nu\lambda} &=& \partial_\mu B_{\nu\lambda} + \partial_\nu B_{\lambda\mu} + 
\partial_\lambda B_{\mu\nu}\, .
\end{eqnarray}
The coupling with dynamical fermionic matter acting
as spectators fields in the duality transformation
has also been considered in \cite{BF}.

The non invariant theory (\ref{PB10}) presents ten primary and four secondary constraints totaling fourteen second-class constraints \cite{hari} leading to three propagating degrees of freedom. On the other hand, the gauge
invariant version has four primary and four secondary constraints of the first class type that are however not independent, forming a reducible system of constraints. After gauge fixing we end up with fourteen second-class constraints as well.  Physically one can see that there is a surviving longitudinal mode coming from the $B_{\mu\nu}$ field. This is the mode that couples to the $A_\mu$ field to produce the massive boson \cite{lahiri}. The Hamiltonian is correspondingly first class. The Hamiltonian equivalence between these two systems has been established in \cite{hari} through the constraint conversion approach.

To study the duality of nonlinear models involving the $B\wedge F$ term in general and the
BI model in particular is the main focus of this paper.
Let us recall that the study of the electric-magnetic duality symmetry
in BI theory, as a non-linear generalization of Hodge
duality was first recognized by Schr\"odinger \cite{S1}, 
and may be viewed as a special case of S-duality.
The inclusion of a $B\wedge F$ term seems natural in this context.
Besides the motivations already mentioned, it was realized sometime ago that the theory admits exact solutions
exhibiting exceptional  properties \cite{BB}.
{}From the String Theory perspective, this relates to the recent interest in open string theory 
in a constant background Kalb-Ramond potential $B_{\mu \nu}$
and thus with gauge theory in a flat non-commutative
space-time \cite{seiwit}.

   Led by the equivalence (\ref{PB10}) $\rightleftharpoons$ (\ref{BI40}) in the linear case, we ask ourselves if 
the duality equivalence can be extended in an arbitrary way. In particular, given a 
``general'' nonlinear self dual model $NSD_{B \wedge F}$,
\begin{equation} \mathcal{L}_{NSD_{B \wedge F}}=\mathcal{F}\left( A_\mu A^\mu , B_{\mu \nu}
 B^{\mu \nu}\right)- \frac{\chi\theta}{2}
\varepsilon^{\mu \nu \rho \lambda }
\partial_\mu B_{\nu \rho} A_\lambda \;\; , 
\end{equation}
we want to know what is the corresponding topologically massive dual equivalent.

To answer this question, in Section II we use the auxiliary field 
technique to linearize the $NSD_{B \wedge F}$  model
in terms of the arguments $A^{2} = A_{\mu}A^{\mu}$  and $B^{2} = B_{\mu \nu}B^{\mu \nu}$
and employ the iterative embedding procedure \cite{IW, AJRRW} to construct a
gauge invariant theory out of the $NSD_{B \wedge F}$.
This procedure, as appropriate for a gauge embedding algorithm,
produces changes in the nature of the constraints of the SD theory.
However, instead of focusing on the constraints, we iteratively introduce 
counter-terms into the Lagrangian density built with powers of the SD Euler vectors and tensors \cite{IW, AJRRW}. As 
discussed in these references, the resulting theory is on-shell equivalent with the original
nonlinear SD model but is, by construction, bound to be gauge invariant.  
To illustrate this procedure a few examples are developed at the end of the section.
It is important to mention,
at this juncture, that since the counter-terms added to make the theory gauge invariant should vanish on-shell
in order to preserve the dynamical contents of the original model, the resulting equivalence in the quantum regime can not, in general, be warranted on the basis of the present analysis.  The possibility that the equivalence is preserved after quantization must be examined in individual basis and is beyond the scope of the present investigation.

In section 3, we specialize to the case when the electrodynamics theory is Born-Infeld.
The inclusion of dynamical matter coupled minimally to both $A_\mu$ and $B_{\mu\nu}$ is discussed at
the end of the section. Our results and perspectives are discussed in our final Section IV.

\section{Generalized Gauge Embedding}

We will follow the notation and procedure outlined in Ref.\cite{BF}. The restricted case
where the nonlinearity is confined to the vector potential sector only will be dealt with
first. Latter on we shall
extend the  nonlinearity to the Kalb-Ramond potential as well. In this sense, the theory to
 be studied first in this section has the following form

\be
{\mathcal{L}_{SD}}=\beta^2 \; g\left(m^2 A^2/2\beta^2 \right) - \frac{1}{4} B_{\mu
\nu}B^{\mu \nu} - \frac{\chi\theta}{2} \varepsilon^{\mu \nu \rho \lambda
} 
\partial_\mu B_{\nu \rho} A_\lambda \;\; .
\ee
The presence of the dimensional control parameter $\beta$ is two fold. It provides the correct
canonical dimension and reproduce the linear case (\ref{PB10}) in the limit $\beta\to\infty$
(in which case it is convenient to normalize the nonlinear function such that $g'(0)=1$). Other
cases will be studied below for illustrative purposes.
To disclose the inherent self-duality of this action it is interesting to compute the equation
of motion for the fields,
\begin{eqnarray}
A^\alpha&=&- \frac{\chi\theta}{2 m^2 g'} \varepsilon^{\alpha \mu \nu \rho}
\partial_\mu B_{\nu \rho}\;\; , \nonumber\\
 B^{\mu \nu}&=&\chi\theta  \varepsilon^{\mu \nu \rho \lambda} \partial_\rho
A_\lambda
\label{eqm2}\;\; ,
\end{eqnarray}
where prime means derivative with respect to the argument. Notice that 
\begin{eqnarray}
\partial_\alpha A^\alpha&=&- \frac{\chi\theta}{2 m^2 } \partial_\alpha\left(\frac 1{g'}\right)
\varepsilon^{\alpha \mu \nu \rho} \partial_\mu B_{\nu \rho}\;\; , \nonumber\\
\partial_\mu B^{\mu \nu}&=& \partial_\nu B^{\mu \nu} = 0\;\; .
\label{eqmm2}
\end{eqnarray}
Further algebra leads to,
\begin{eqnarray}
\label{SS10}
 (\square + \frac{ m^2}{\theta^2} g')A^\mu &=& \partial^\mu(\partial_\alpha A^\alpha)\;\; , \nonumber\\
 (\square + \frac{ m^2}{\theta^2} g')B^{\mu  \nu} &=& -g' \partial_\rho \left(\frac{1}{g'}\right) H^{\mu \nu
 \rho}\;\; .
\end{eqnarray}
It is noticeable that although the nonlinearity is initially allocated in the vector potential sector,
the equations of motion of both sectors displays their presence due to the coupling provided by the $B\wedge F$ term
and decouple in the linear limit.

We shall define the duality operation in the derivative sense \cite{BF}. By a simple index
counting argument we find that the duality $(*)$-operation maps $A_\mu$ into $B_{\mu \nu}$ and vice-versa, 
\begin{eqnarray}
^*A^\alpha &\equiv& - \frac{\theta}{2 m^2 g'} \varepsilon^{\alpha \mu \nu
\rho}
\partial_\mu B_{\nu \rho}\;\; , \nonumber\\
^*B^{\mu \nu}&\equiv& \theta\varepsilon^{\mu \nu \rho \lambda}
\partial_\rho A_\lambda\;\; . \label{d2}
\end{eqnarray}
Upon use of the equations of motion (\ref{eqm2}), we prove the double duality property, $*\cdot * = 1$
of the $(*)$-operation (\ref{d2}), 
\begin{eqnarray}
^*(^*A^\mu) &=& A^\mu\;\; , \nonumber\\
 ^*(^*B^{\mu \nu})&=&B^{\mu \nu}\;\; , \label{autodual2}
\end{eqnarray}
which allows for consistent self and anti-self dual solutions. 

To apply the gauge embedding method we need to linearize the function $g(x)$ in terms of the argument,
 which can be realized by the auxiliary field technique,
\begin{equation}
\label{DD10}
g\left(x\right) \rightarrow \frac{x}{\lambda} + f(\lambda)\;\; ,
\end{equation}
which is basically a Legendre transformation.
The exact form of $f(\lambda)$ for arbitrary $g(x)$, found in \cite{IW}, is given by the usual Legendre
transformation algorithm.  Taking variations with respect to $x$ in (\ref{DD10}) allows us to write $x=x(\lambda)$
\begin{eqnarray}
\label{DD20}
g'(x) = \frac 1\lambda \;\;\;\Rightarrow\;\;\; x=x(\lambda)\;\; ,
\end{eqnarray}
while variations with respect to $\lambda$ gives
\begin{eqnarray}
\label{DD30}
f'(\lambda) = \frac x{\lambda^2}\;\; ,
\end{eqnarray}
where prime in both cases has the meaning of derivative with respect to the argument. Integrating (\ref{DD30}) and using (\ref{DD20}) gives us the desired result
\begin{equation}
\label{DD40}
f(\lambda)=\int^\lambda d\sigma \frac{1}{\sigma^2}
\left[\bar{g'}(\sigma^{-1})\right]\;\; ,
\end{equation}
where the bar over the function indicates its functional inverse in the sense $\bar h(h(x))= x$.
Once the form of the function $f(\lambda)$ is found, we may return to the discussion of the gauge embedding.
Rewriting the linearized  $NSD_{B\wedge F}$ Lagrangian as,
\begin{eqnarray}
{\mathcal{L}_\lambda}= \frac{m^2 A^2}{2\lambda} + \beta^2 f(\lambda) -
\frac{1}{4} B_{\mu \nu}B^{\mu \nu} - \frac{\chi\theta}{2}
\varepsilon^{\mu \nu \rho \lambda }
\partial_\mu B_{\nu \rho} A_\lambda\;\; , \label{Llamb}
\end{eqnarray}
allows us to compute the Euler tensors through the variations of ${\cal L}_\lambda$ as 
\be
\delta {\cal L}_\lambda = K_\mu \delta A^\mu + M_{\mu \nu} \delta B^{\mu \nu}\;\; ,
\ee
where $K_\mu$ and $M_{\mu \nu}$ are 
\begin{eqnarray}
\label{KnL}
K^\sigma &=& \frac{m^2 A^\sigma}{\lambda} - \frac{\chi\theta}{2}
\varepsilon^{\mu \nu \rho \sigma} \partial_\mu B_{\nu \rho}\;\; , \\\nonumber
M^{\nu \rho}&=&-\frac{1}{2}B^{\nu \rho} + \frac{\chi\theta}{2}
\varepsilon^{\mu \nu \rho \lambda} \partial_\mu A_\lambda\;\; .
\end{eqnarray}
Following the same steps as in the linear case we find, after some algebra, the linear dual as
\begin{equation}
\label{L}
{\mbox{}^* \mathcal{L}}_{\lambda}=\mathcal{L}_\lambda
- \frac{\lambda}{2 m^2} K^\mu K_\mu +M^{\mu \nu}M_{\mu \nu}\;\; .
\end{equation}
Substituting Eqs.(\ref{KnL})  into (\ref{L}) we get the following Lagrangian density,
\begin{eqnarray}
{\mbox{}^* \mathcal{L}}_{\lambda}=\beta^2 f(\lambda)-\frac{1}{4}F_{\mu \nu} F^{\mu \nu} +
\frac{\lambda}{24 m^2} H_{\mu \nu \rho} H^{\mu \nu \rho} -
\frac{\chi}{2\theta} \varepsilon^{\mu \nu \rho \lambda }
 B_{\nu \rho} \partial_\mu A_\lambda \;\; ,
\end{eqnarray}
which is still dependent on the $\lambda$-field.
The subsequent elimination of the auxiliary linearizing variable $\lambda$, that can be done in a systematic way,
leads, in general, to a nonlinear structure for the field strength of the Kalb-Ramond field, not the Maxwell vector
field, as it would be  naively expected. It should be noticed that to eliminate the auxiliary variable $\lambda$,
may or may not be simple, technically speaking, in the sense that a solution in terms of elementary functions may not
be possible.  More on this subject below. It is also important to notice that this interplay between the Maxwell
 and KR sectors, the inversion of the coupling constant $\theta \to 1/\theta$,  as well as the disclosing of the
symmetry hidden in the SD-representation are the main features of the linearization/embedding procedure associated
to the $(*)$-operation defined in 
(\ref{d2}).  The exact
 structure will certainly depend on each particular case. The linearization/embedding method is applied next to
some simple examples in order to illustrate these features.

An interesting model displaying the properties studied above presents a logarithmic nonlinearity.
A logarithmic $U(1)$ gauge theory has been investigated \cite{soleng} as an example of the class of theories constructed
in \cite{Altshuler} to discuss inflation.
Its Lagrangian density is defined as
\begin{equation}
{\mathcal{L}}= \beta^2 \ln \left(1+ \frac{m^2 A^2}{2\beta^2} \right) -\frac{1}{4} B_{\mu
\nu}B^{\mu \nu} - \frac{\chi\theta}{2} \varepsilon^{\mu \nu \rho \lambda
}
\partial_\mu B_{\nu \rho} A_\lambda\;\; , \label{Log}
\end{equation}
whose large $\beta$ limit gives back the linear case.
While this particular theory appears to have no direct relation to the brane-theory,
it serves as a toy-model illustrating that
certain non-linear field theories can produce
particle-like solutions realizing the limiting curvature hypothesis \cite{soleng} also for gauge fields.

To linearize the function
\begin{equation}
g(x)= \ln \left(1+ x\right) \;\;\; ; \;\;\; x=\frac {m^2 A^2}{2\beta^2}\;\; ,
\end{equation}
we use formula (\ref{DD40}) to obtain the auxiliary function
\begin{equation}
f(\lambda) = \ln \lambda + \frac 1\lambda\; .
\end{equation}
The linearized Lagrangian density then becomes
\begin{equation}
{\cal L}_{\lambda} = \beta^2 f(\lambda) + \frac {m^2}{2\lambda} A_\mu A^\mu - \frac 14 B_{\mu\nu} B^{\mu\nu}
+\frac {\chi\,\theta}4 \epsilon^{\mu\nu\lambda\rho} B_{\mu\nu} F_{\lambda\rho}\; .
\end{equation}
The embedding is now easily performed and gives the linearized dual
\begin{equation}
\label{PP10}
{\mathcal{\mbox{}^* L}_{\lambda}}=-\frac{1}{4}F_{\mu \nu} F^{\mu \nu}+\beta^2 \left( \ln \lambda + \frac 1\lambda \right)
+ \frac{\lambda}{12 m^2} H_{\mu \nu \rho} H^{\mu \nu \rho} -
\frac{\chi}{2\theta} \varepsilon^{\mu \nu \rho \lambda }
 B_{\nu \rho} \partial_\mu A_\lambda\, .
\end{equation}
To obtain the effective dual action we need to solve for the auxiliary field $\lambda$,
\begin{equation}
\lambda=\frac 1{y^2} \left(-1 + \sqrt{1\, +\, 2\, y^2}\right) \;\;\; ; \;\;\; y^2 = \frac {H^2}{6 m^2 \beta^2}\;\; ,
\end{equation}
which, upon substitution back on (\ref{PP10}), produces the gauge invariant dual Lagrangian density,
\begin{equation}
{\mathcal{\mbox{}^*L}_{eff}}=-\frac{1}{4}F_{\mu \nu} F^{\mu \nu}+ 
\beta^2 \sqrt{1 \, + \, 2\, y^2} + \beta^2 \ln{\left( \frac {-1 + \sqrt{1 \, + \, 2\, y^2}}{y^2}\right)} 
- \frac{\chi}{2\theta}
\varepsilon^{\mu \nu \rho \lambda }
 B_{\nu \rho} \partial_\mu A_\lambda\;\; .
\end{equation}
As argued, the nonlinearity has been swapped to the KR sector and $\theta \to 1/\theta$. To finish, let us examine the limit of large $\beta$. Indeed
\begin{equation}
\beta^2\left[\sqrt{1 \, + \, 2\, y^2} + \ln{\left( \frac {-1 + \sqrt{1 \, + \, 2\, y^2}}{y^2}\right)}\right]\to \beta^2 \frac {y^2}{2} = \frac 1{12 m^2} H_{\mu \nu \rho} H^{\mu \nu \rho}\; ,
\end{equation}
as expected. A disconnected $\beta^2$ term, that does not contribute dynamically to the equation of motion,
has been disregarded.

Another interesting nonlinear model is the self-dual rational model defined as
\begin{equation}
{\mathcal{L}}= \frac{q}{p} \beta^2
\left(\frac{1}{\beta}A_\mu A^\mu\right)^{\frac{p}{q}}-
\frac{1}{4} B_{\mu \nu}B^{\mu \nu} - \frac{\chi\theta}{2}
\varepsilon^{\mu \nu \rho \lambda }
\partial_\mu B_{\nu \rho} A_\lambda \; ,
\end{equation}
with $p$ and $q$ integers and the limit back into the linear case being $p/q \to 1$. Since we are not interesting in taking the large $\beta$ limit we have chosen $m^2 = 2 \beta$.  This model is interesting particularly when $p/q =\mbox{integer}$ in which case the monomials represent usual self-interactions. It is closely related to the nonperturbative gluondynamics model proposed long ago by Pagels and Tomboulis \cite{PG} which, in its abelian sector, can be reduced to a strongly nonlinear electrodynamics. 
It may also be of interest to study the Bardeen model \cite{bardeen} of black holes coupled to nonlinear electrodynamics
leading to nonsingular metrics \cite{EABG}.
The nonlinear function
\begin{equation}
\label{TT10}
g\left( A^2/\beta \right)=\frac{q}{p} 
\left(\frac{1}{\beta}A_\mu A^\mu\right)^{\frac{p}{q}},
\end{equation}
leads to a well defined dual model for  all $p$ and  $q$ integers with the proviso $p/q \neq \{1, 1/2 \}$. In the
$p=q$ case we return to the usual self dual model, while $q=2p$
is problematic and will not be discussed here.
As mentioned, the normalization of the nonlinear function has been modified in this example since the linear case is not
taken by the limit $\beta\to\infty$. Otherwise we may consider the function $g \sim (1 + m^2A^2/ 2\beta)^{p/q}$. This
modification leads to a well defined linearized dual action. The solution for $\lambda = \lambda(H^2)$ cannot however
be written in terms of elementary functions so that we will not pursue this example any further.

To linearize the rational function in (\ref{TT10}) we use the auxiliary function
\begin{equation}
f(\lambda)= \frac{q-p}{p} \lambda^\frac{p}{q-p}\; .
\end{equation}
Following the duality procedure we find the effective dual Lagrangian as
\begin{equation}
{\mbox{}^*\mathcal{L}}_{eff}=-\frac{1}{4}F_{\mu \nu} F^{\mu \nu} +
 \frac{q-2p}{p}(\beta^2) \left(- \frac{H^2}{24\beta^3} \right)^\frac{p}{2p-q} - \frac{\chi}{2\theta}
\varepsilon^{\mu \nu \rho \lambda }
 B_{\nu \rho} \partial_\mu A_\lambda\; .
\end{equation}
One can check that the limit $p=q$ gives us back the usual mapping for the SD model into the topologically massive $B\wedge F$ model with $\beta$ playing the role of the mass of the elementary excitations.

After this preliminary analysis on the structure of the duality transformation with the nonlinearity
confined in Maxwell $A_\mu$-sector, which displays the main features of the procedure, let us next consider some more general situations. First we consider the following Lagrangian density
\begin{equation}
{\mathcal{L}_{SD}}=\beta^2 g\left( m^2 A^2/2\beta^2 \right) - \frac {\beta^2}2 h\left( m^2 B^2/2\beta^3\right)-
\frac{\chi\theta}{2}
\varepsilon^{\mu \nu \rho
\lambda }
\partial_\mu B_{\nu \rho} A_\lambda\; .
\end{equation}
The  equations  for motion of two fields are,
\begin{eqnarray}
A^\alpha&=&- \frac{\chi\theta}{2 m^2 g'} \varepsilon^{\alpha \mu \nu \rho}
\partial_\mu B_{\nu \rho}\; , \nonumber \\
B^{\mu \nu}&=&\beta \frac{\chi\theta}{ m^2 h'}  \varepsilon^{\mu \nu \rho \lambda}
\partial_\rho A_\lambda\; .
\end{eqnarray}
{}From these equations the following relations are found,
\begin{eqnarray}
\partial_\mu A^\mu&=& - \frac{\chi\theta}{2 m^2}  \varepsilon^{\mu \nu \rho \lambda}
\partial_\mu \left(\frac{1}{g'} \right) \partial_\nu B_{\rho \lambda}\; , \nonumber \\
\partial_\mu B^{\mu \nu}&=& \beta \frac{\chi\theta}{m^2} \varepsilon^{\mu \nu \rho \lambda}
\partial_\mu \left(\frac{1}{h'} \right) \partial_\rho A_{\lambda}\; ,
\end{eqnarray}
from where the radiative equations follows
\begin{eqnarray}
 (\square + \frac{m^4}{\beta \theta^2} g'h')A^\mu &=& h' \partial_\nu \left(\frac{1}{h'}
\right)F^{\mu \nu} + \partial^\mu (\partial_\nu A^\nu)\; , \nonumber\\
 (\square + \frac{m^4}{\beta\theta^2} g'h')B^{\mu  \nu} &=& -g' \partial_\rho \left(\frac{1}{g'}\right) H^{\mu \nu
 \rho}+ \partial_\rho (\partial^\mu B^{\nu \rho} + \partial^\nu B^{\rho \mu})\; .
\end{eqnarray}
Note that in this nonlinear case the fields will have longitudinal components, which is the usual behavior of real materials. In the linear limit where $g'=h'=1$ we return to the transverse propagation given by (\ref{PB10}).

Next we define the dual operation in the usual way as
\begin{eqnarray}
^*A^\alpha &\equiv& - \frac{\theta}{2 m^2 g'} \varepsilon^{\alpha \mu \nu \rho}
\partial_\mu B_{\nu \rho} \; ,\nonumber \\
^*B^{\mu \nu} &\equiv& \beta \frac{\theta}{ m^2 h'}  \varepsilon^{\mu \nu \rho \lambda}
\partial_\rho A_\lambda\; ,
\end{eqnarray}
so that, on-shell, the relations (\ref{autodual2}),
will validate the definition of dual fields. If the relations are combined, we conclude that
\begin{eqnarray}
^*A_\mu &=& \chi A_\mu \; ,\nonumber \\
 ^*B_{\mu \nu} &=& \chi B_{\mu \nu}\; ,
\end{eqnarray}
depending of the  $\chi$ signal, the theory  corresponds the self dual model or anti-selfdual model.

The linearization  of both $g(A^2)$ and $h(B^2)$ functions follows the same steps of the previous sections,
\begin{equation}
h\left(x \right) \rightarrow \frac{x}{\kappa} + l(\kappa)\; ,
\end{equation}
etc. In terms of the auxiliary fields, the nonlinear model becomes linearized as
\begin{equation}
{\mathcal{L}_{\lambda,\kappa}}= \frac{m^2 A^2}{2\lambda} -
\frac{m^2 B^2}{4\beta\kappa} + \beta^2 f(\lambda) -\frac{ \beta^2}{2} l(\kappa) - \frac{\chi\theta}{2} \varepsilon^{\mu
\nu \rho \lambda }
\partial_\mu B_{\nu \rho} A_\lambda \label{Llamb2}\; .
\end{equation}
Now the stage is set for dualization. Using the gauge embedding procedure we find the dual equivalent linearized
Lagrangian density
\begin{eqnarray}
{\mbox{}^*\mathcal{L}_{\lambda , \kappa}}=\beta^2 f(\lambda)  -\frac{ \beta^2}{2} l(\kappa) - \frac{\beta\kappa}{4 m^2}F_{\mu\nu} F^{\mu \nu} + \frac{\lambda}{12 m^2} H_{\mu \nu \rho} H^{\mu \nu\rho} - \frac{\chi}{2\theta} \varepsilon^{\mu \nu \rho \lambda }
 B_{\nu \rho} \partial_\mu A_\lambda \; .
\end{eqnarray}
Solving for both auxiliary fields, the inherent nonlinearity is recovered, yielding
\begin{eqnarray}
{\mbox{}^*\mathcal{L}_{TM}}= - \beta^2 \mathcal{F}( F^2) +
\beta^2 \mathcal{H}(H^2) - \frac{\chi}{2\theta}
\varepsilon^{\mu \nu \rho \lambda }
 B_{\nu \rho} \partial_\mu A_\lambda\; .
\end{eqnarray}

Finally, we end up this discussion on the general setting, considering a situation where the Maxwell and the
KR fields are taken on equal footing,
\begin{eqnarray}  \mathcal{L}_{SD_{B \wedge F}}=\beta^2 g\left( A^2 / \beta - B^2 /\beta^2 \right)- \frac{\chi\theta}{2}
\varepsilon^{\mu \nu \rho \lambda }
\partial_\mu B_{\nu \rho} A_\lambda \nonumber\; .
\end{eqnarray}
{}Following the previous procedure we obtain
\begin{eqnarray}
\mbox{}^*\mathcal{L}_{\lambda}=\lambda\left(\frac {-1}{16} F_{\mu\nu}F^{\mu\nu} + \frac 1{24\beta}
H_{\mu\nu\rho}H^{\mu\nu\rho}\right) 
+ \beta^2 f(\lambda)- \frac \chi{2\theta} A_\mu \varepsilon^{\mu\nu\rho\sigma}\partial_\nu B_{\rho\sigma}\; ,
\end{eqnarray}
where $f(\lambda)$ is the linearizing function. The elimination of the $\lambda$ field yields the desired result for the nonlinear topologically massive version as
\begin{equation}
{\mbox{}^*\cal L}_{TM} = \beta^2 {\cal F}\left( \frac {-1}{16} F_{\mu\nu}F^{\mu\nu} + \frac 1{24\beta}
H_{\mu\nu\rho}H^{\mu\nu\rho}
\right) - \frac{\chi}{2\theta} \varepsilon^{\mu \nu \rho \lambda }
 B_{\nu \rho} \partial_\mu A_\lambda\; .
\end{equation}
An illustration of this last case will be presented in the following section, in the context of the BI theory,
that is our main topic of interest here.

\section{Born-Infeld Nonlinear Electrodynamics}

Born--Infeld is an amazing theory.  This is a  non-linear electrodynamics  created upon the desire to find a non-singular
field theory, i.e.,  a model whose action would lead to a  bounded field strength. Alternatively,
Euler and Heisenberg \cite{HE}
discovered that vacuum polarization
effects can be simulated classically
by a non-linear theory. Also, as discussed in the Introduction, in string
theory one has found effective actions describing
non-linear electromagnetism \cite{RMT85}.

In this section we study the dual correspondence between some nonlinear $SD_{B\wedge F}$ model \cite{khare}
and Born-Infeld-BF model (BIBF)
employing the gauge embedding procedure. The Lagrangian
density for the non invariant model has been proposed in \cite{khare} in an investigation of duality in $D=3$ Chern-Simons theory and reads
\begin{equation}
\mathcal{L}= \beta^2 \sqrt{1+\frac{m^2}{\beta^2}A_\mu A^\mu}-
\frac{1}{4} B_{\mu \nu}B^{\mu \nu} - \frac{\chi\theta}{2}
\varepsilon^{\mu \nu \rho \lambda }
\partial_\mu B_{\nu \rho} A_\lambda\; .
\end{equation}
Here we have restored the two dimensional parameters.
The $\beta$ is a parameter inserted for dimensional reasons and in the limit $\beta \rightarrow \infty$
gives back the usual  $SD_{B \wedge F}$ 
model after discarding a dynamically unimportant constant and $m$ assumes its usual interpretation as the mass of the excitations. 

Using the procedure developed in the preceding section we get
\begin{equation}
\label{PP34}
f(\lambda)=\left(\frac{\lambda}{4}  +
\frac{1}{\lambda} \right)\; ,
\end{equation}
such that the dual linearized Lagrangian density, after the implementation of the embedding procedure, becomes 
\begin{equation}
{\mbox{}^*\mathcal{L}}_{\lambda}=-\frac{1}{4}F_{\mu \nu} F^{\mu \nu}+\beta^2
\left(\frac{\lambda}{4}  + \frac{1}{\lambda} \right) +
\frac{\lambda}{24 m^2} H_{\mu \nu \rho} H^{\mu \nu \rho} -
\frac{\chi}{2\theta} \varepsilon^{\mu \nu \rho \lambda }
 B_{\nu \rho} \partial_\mu A_\lambda\; ,
\end{equation}
with the proviso that the fields have been scaled as $A_\mu\to A_\mu/\theta$ and $B_{\mu\nu}\to B_{\mu\nu}/\theta$.
Solving for the auxiliary field $\lambda$ gives
\begin{equation}
\lambda=\left(\frac{1 }{24 m^2\beta^2} H^2
+\frac{1}{4}\right)^{-\frac{1}{2}}\; ,
\end{equation}
which, upon substitution back into the linearized dual, produces the gauge invariant topologically massive $BIBF$ model
\begin{equation}
{\mbox{}^*\mathcal{L}}_{eff}=-\frac{1}{4}F_{\mu \nu} F^{\mu \nu} + \beta^2
\sqrt{1+\frac{1}{6 m^2\beta^2}H_{\mu \nu \rho} H^{\mu \nu \rho}} -
\frac{\chi}{2\theta} \varepsilon^{\mu \nu \rho \lambda }
 B_{\nu \rho} \partial_\mu A_\lambda\; .
\end{equation}
Notice that in the limit $\beta \rightarrow \infty$ we recover the usual $TM_{B \wedge F}$ 
model, as discussed in \cite{BF}. Notice also the swapping of the nonlinearity from the Maxwell to the Kalb-Ramond
sector.

Let us consider next the situation where the nonlinearity is present in both sectors.
To illustrate these cases we consider first the  Born-Infeld-Log
model,
\begin{eqnarray}
\mathcal{L}= -\beta^2 \ln \left(\frac{1}{\beta}A_\mu A^\mu
\right)+ \alpha^2 \sqrt{1 - \frac{1}{2\alpha^2}B_{\mu \nu}B^{\mu
\nu}} - \frac{\chi\theta}{2} \varepsilon^{\mu \nu \rho \lambda }
\partial_\mu B_{\nu \rho} A_\lambda\; ,
\end{eqnarray}
with $\alpha$ and $\beta$ with mass dimension two.  After some algebra we find the dual theory to be
\begin{eqnarray}
\mbox{}^*\mathcal{L}_{eff}= \alpha^2 \sqrt{1-\frac{1}{2\alpha^2}F_{\mu
\nu} F^{\mu \nu}}+\beta^2 \ln \left(\frac{H^2}{24 \beta^3} \right)
- \frac{\chi}{2\theta} \varepsilon^{\mu \nu \rho \lambda }
 B_{\nu \rho} \partial_\mu A_\lambda\; .
\end{eqnarray}

To exemplify the second situation we consider a generalization of the model presented in \cite{khare}
\begin{eqnarray} \mathcal{L}_{SD_{B \wedge F}}= \beta^2 \sqrt{1+\frac{m^2}{\beta^2} A_\mu A^\mu
-\frac{1}{2\beta^2}B_{\mu \nu} B^{\mu \nu}}- \frac{\chi\theta}{2}
\varepsilon^{\mu \nu \rho \lambda }
\partial_\mu B_{\nu \rho} A_\lambda\; ,
\end{eqnarray}
which is linearized by Eq.(\ref{PP34}) and leads after embedding and elimination of the auxiliary field to
\begin{equation}
{\cal L}_{TM} = \beta^2 \sqrt {1 + \frac{1}{6m^2 \beta^2} H^{\mu
\nu \rho}H_{\mu \nu \rho}-\frac{1}{2 \beta^2} F^{\mu \nu} F_{\mu
\nu} } - \frac{\chi}{2\theta} \varepsilon^{\mu \nu \rho \lambda }
 B_{\nu \rho} \partial_\mu A_\lambda\; .
\end{equation}

To conclude this section we want consider the coupling to dynamical fermionic matter, coupled both to the Maxwell and the KR fields and discuss the duality transformation.  To be specific let us analyze the following Lagrangian density with minimal coupling in both tensorial sectors,
\begin{eqnarray}
\mathcal{L}_{SD_{B \wedge F}}&=&\beta^2 g\left(\frac {m^2 A^2}{2 \beta^2} \right) - \frac {\beta^2}2
h\left( \frac{m^2 B^2}{2 \beta^3}\right)- \frac{\chi\theta}{2}
\varepsilon^{\mu \nu \rho \lambda }
\partial_\mu B_{\nu \rho} A_\lambda \nonumber\\
&+& {\cal L}_D- e A^\mu  J_\mu - \frac gm  B_{\mu \nu} {\cal J}^{\mu \nu} \; ,
\end{eqnarray}
where $e$ and $g$ are the strengths of the vector and tensor couplings and
\begin{equation}
\label{45}
{\cal L}_{D} =  \bar{\psi}(i\partial\!\!\! /  -M)\psi \; ,
\end{equation}
is the Dirac Lagrangian density and
the rank-1 and rank-2 fermionic currents are
\begin{eqnarray}
\label{100}
J^\mu = \bar{\psi}\gamma^\mu \psi\; ,
\end{eqnarray}
and
\begin{eqnarray}
\label{399}
{\cal J^{\mu\nu}} = {\cal C} \bar\psi \gamma^\mu\gamma^\nu\psi\; ,
\end{eqnarray}
where ${\cal C}$ is a complex normalization constant.

After linearization we follow previous procedure: compute the Euler tensors
\begin{eqnarray}
\label{WW27}
\label{5001}  K_\mu^D &=& \frac{m^2 A^\sigma}{\lambda} - \frac{\chi\theta}{2}
\varepsilon^{\mu \nu \rho \sigma} \partial_\mu B_{\nu \rho} - e J_\mu \; ,\nonumber\\
 M_{\mu\nu}^D &=&-\frac{m^2}{2\kappa\beta}B^{\nu \rho} + \frac{\chi\theta}{2}
\varepsilon^{\mu \nu \rho \lambda} \partial_\mu A_\lambda - \frac gm  {\cal J}_{\mu\nu}\; ,
\end{eqnarray}
from where the linearized dual action is obtained,
\begin{eqnarray}
\mbox{}^*{\cal L}_{\lambda,\kappa}&=& \mathcal{L}_{SD_{B \wedge F}} - \frac \lambda{2 m^2}
K_D^2 + \frac {\kappa\beta}{m^2} M_D^2\nonumber\\
&=& \beta^2 f(\lambda) -\frac {\beta^2}2 l(\kappa) -  \frac{\chi}{2\theta}B_{\nu\rho}\varepsilon^{\mu \nu \rho \lambda}
\partial_\mu A_\lambda \nonumber\\
&-& \frac \lambda{24 m^2} H^2 - \frac {\lambda \; e^2}{2 m^2} J_\mu J^\mu + \frac {\lambda\chi\; e}{2 m^2} J_\mu \varepsilon^{\mu\nu\rho\sigma}\partial_\nu B_{\rho\sigma}\nonumber\\
&-& \frac {\kappa\beta}{4 m^2}F_{\mu\nu}F^{\mu\nu} + \frac{\kappa\beta\; g^2}{m^4} {\cal J}_{\mu\nu} {\cal J}^{\mu\nu}
- \frac {\kappa\beta \chi \; g}{m^3}  {\cal J}_{\mu\nu} \varepsilon^{\mu\nu\rho\sigma}\partial_\rho 
A_{\sigma}\; ,
\end{eqnarray}
after the expressions for the currents (\ref{WW27}) are substituted back into the action. Elimination of the auxiliary fields
in the linear dual action produces the full topologically massive dual action
\begin{eqnarray}
&&{\mbox{}^*\cal L}_{TM}=\beta^2{\cal H} \left(\frac{1}{24m^2} H^{\mu \nu \rho}
H_{\mu \nu \rho} -\frac {e\;\chi}{2 m^2} \, \varepsilon^{\mu \nu \rho \lambda}
\partial_\mu B_{\nu \rho} J_\lambda -\frac{e^2}{2 m^2} J^\mu J_\mu
\right)+{\cal L}_D \nonumber \\ & &+\,\beta^2{\cal F}\left(-\frac{\beta}{4m^2}
F^{\mu \nu} F_{\mu \nu} -\frac {\beta\chi\; g}{m^3} \, \varepsilon^{\mu\nu \rho \sigma}
\partial_\rho A_{\sigma} {\cal J}_{\mu \nu} -
\frac{\beta \; g^2}{m^4} {\cal J}^{\mu \nu} {\cal J}_{\mu \nu}
\right)+\frac \chi{2\theta} \, \varepsilon^{\mu \nu \rho \lambda}
\partial_\mu B_{\nu \rho} A_\lambda\; .
\end{eqnarray}
The direct application of this formulae to the BI--Log model, including fermionic matter, leads to the following
topologically massive action,
\begin{eqnarray}
{\cal L}_{eff}&=&\beta^2 \ln \left(\frac{1}{24m^2} H^{\mu \nu \rho}
H_{\mu \nu \rho} -\frac {e\;\chi}{2 m^2} \, \varepsilon^{\mu \nu \rho \lambda}
\partial_\mu B_{\nu \rho} J_\lambda -\frac{e^2}{2 m^2} J^\mu J_\mu\right)+ \frac \chi
{2\theta} \, \varepsilon^{\mu \nu \rho \lambda}
\partial_\mu B_{\nu \rho} A_\lambda \nonumber \\ &+&{\cal L}_D+ \,\beta^2\sqrt{-\frac{\beta}{4m^2}
F^{\mu \nu} F_{\mu \nu} -\frac {\beta\chi\; g}{m^3} \, \varepsilon^{\mu\nu \rho \sigma}
\partial_\rho A_{\sigma} {\cal J}_{\mu \nu} -
\frac{\beta \; g^2}{m^4} {\cal J}^{\mu \nu} {\cal J}_{\mu \nu}}\; .
\end{eqnarray}
It is quite interesting to see that duality mapping from SD--$B\wedge F$ to the  TM--$B\wedge F$ model transforms a minimal coupling into a magnetic--like, non minimal coupling of the matter with the tensors participating in the dualization.  Although matter is spectator in the whole process of duality, it is amazing to see the appearance of Thirring--like terms in the dual theory, not present in the original model.  This happens to maintain unaltered the fermionic dynamics before and after dualization of the tensorial fields.

\section{conclusions}

In this paper we studied the dual equivalence between the nonlinear 
generalization of the self dual and the topologically massive $B\wedge F$ models 
in 3+1 dimensions. We have used the formalism of Noether embedding, 
which provides a clear physical meaning of the
duality equivalence since the counter-terms that are added to provide the gauge symmetry
vanish on-shell. In this paper we deal specifically with the nonlinear case.
This is accomplished by linearizing the nonlinear 
terms of $A^2$ and $B^2$ by means of a auxiliary field, which can be eliminated later on
to restore the full nonlinearity of the NSD and the generalized TM
models.  
The usual SD-TM dual equivalence is naturally contained in these results as well as the disclosing of the
hidden symmetries of the SD sector which happens in the nonlinear situation as well.
To include the couplings with dynamical matter is a simple operation since in the gauge
embedding procedure the matter fields are just spectators in the dual operation involving the gauge tensors.
Some examples are discussed that both clarify the technique and prove the power of the gauge embedding approach to 
deal with duality equivalence. The main features obtained are inversion of the coupling
constant, which is a usual feature of the S-duality, and the swapping between the Maxwell
and the Kalb-Ramond sectors.  This swapping persists if the coupling to external currents
are included. Also characteristic of the duality mapping involving matter is the appearance of a Thirring
like term and the change of a minimal coupling into a nonminimal, magnetic like, coupling that happens to
preserves the dynamics in the matter sector.

\begin{acknowledgments}
This work is partially supported by CNPq, CAPES, FAPERJ and
FUJB, Brazilian Research Agencies. 
\end{acknowledgments}

\end{document}